\documentclass[]{jfm}

\usepackage{graphicx}
\usepackage{natbib}
\usepackage{subfigure}
\usepackage{color}

\ifCUPmtlplainloaded \else
  \checkfont{eurm10}
  \iffontfound
    \IfFileExists{upmath.sty}
      {\typeout{^^JFound AMS Euler Roman fonts on the system,
                   using the 'upmath' package.^^J}%
       \usepackage{upmath}}
      {\typeout{^^JFound AMS Euler Roman fonts on the system, but you
                   dont seem to have the}%
       \typeout{'upmath' package installed. JFM.cls can take advantage
                 of these fonts,^^Jif you use 'upmath' package.^^J}%
      }
  \else
  \fi
\fi

\ifCUPmtlplainloaded \else
  \checkfont{msam10}
  \iffontfound
    \IfFileExists{amssymb.sty}
      {\typeout{^^JFound AMS Symbol fonts on the system, using the
                'amssymb' package.^^J}%
       \usepackage{amssymb}%
         
       \let\ge=\geqslant  
      }{}
  \fi
\fi

\ifCUPmtlplainloaded \else
  \IfFileExists{amsbsy.sty}
    {\typeout{^^JFound the 'amsbsy' package on the system, using it.^^J}%
     \usepackage{amsbsy}}
    {\providecommand\boldsymbol[1]{\mbox{\boldmath $##1$}}}
\fi

\newcommand\Rey{\mbox{\textit{Re}}}  % Reynolds number
\newcommand\Pra{\mbox{\textit{Pr}}}  % Prandtl number, cf TeX's \Pr product
\newcommand\Ray{\mbox{\textit{Ra}}}  % Rayleigh number
\newcommand\Nus{\mbox{\textit{Nu}}}  % Nusselt number
\newcommand\Lew{\mbox{\textit{Le}}}  % Lewis number
\newcommand\Sch{\mbox{\textit{Sc}}}  % Schmidt number
\newsavebox{\astrutbox}
\sbox{\astrutbox}{\rule[-5pt]{0pt}{20pt}}

\newcommand\eg{e.g.\ }
\newcommand\ie{i.e.\ }

\title[]{Salinity transfer in bounded double diffusive convection}

\author[]%
{Yantao Yang$^1$\thanks{Email address for correspondence: yantao.yang@utwente.nl}, 
 Erwin P. van der Poel$^1$, Rodolfo Ostilla-M\'{o}nico$^1$, Chao Sun$^1$, 
 Roberto Verzicco$^{1,2}$, Siegfried Grossmann$^3$ and Detlef Lohse$^1$}

\affiliation{$^1$Physics of Fluids Group, Faculty of Science and Technology,
                 MESA+ Research Institute, \\ and J. M. Burgers Centre for Fluid Dynamics,
                 University of Twente, \\ PO Box 217, 7500 AE Enschede, The Netherlands. \\
             $^2$Dipartimento di Ingegneria Industriale, University of Rome ``Tor Vergata'',  \\
                 Via del Politecnico 1, Roma 00133, Italy\\
             $^3$Fachbereich Physik, Philipps-Universit\"{a}t Marburg, 
                 Renthof 6, D-35032 Marburg, Germany}

\pubyear{}
\volume{}
\pagerange{}
\date{?; revised ?; accepted ?. - To be entered by editorial office}
\begin{document}

\maketitle

\begin{abstract}
The double diffusive convection between two parallel plates is numerically studied for a series of parameters. The flow is driven by the salinity difference and stabilised by the thermal field. Our simulations are directly compared to experiments by Hage and Tilgner (\emph{Phys. Fluids} 22, 076603 (2010)) for several sets of parameters and reasonable agreement is found. This in particular holds for the salinity flux and its dependence on the salinity Rayleigh number. Salt fingers are present in all simulations and extend through the entire height. The thermal Rayleigh number seems to have minor influence on salinity flux but affects the Reynolds number and the morphology of the flow. Next to the numerical calculation, we apply the Grossmann-Lohse theory for Rayleigh-B\'{e}nard flow to the current problem without introducing any new coefficients. The theory successfully predicts the salinity flux both with respect to the scaling and even with respect to the absolute value for the numerical and experimental results.
\end{abstract}


\section{Introduction}

Double Diffusive Convection (DDC) can occur when the fluid density in a system is affected by two components. Often the diffusivities of the two components are very different. DDC is relevant in many natural environments, such as thermal convection with compositional gradients in astrophysics \citep{Spiegel:1972, Rosenblum_etal:AJ:2011}, the thermohaline effects in the horizontal convection~\citep{Hughes_Griffiths:ARFM:2008}, the sedimentation in salt water~\citep{Burns_Meiburg:JFM:2012}, and the double diffusion in oceanography~\citep{Turner:AnnRev:1974,Schmitt:AnnRev:1994, Schmitt_etal:Science:2005}. In oceanography, the sea water density depends on both temperature and salinity. The Prandtl numbers, \ie the ratio of viscosity to diffusivity of each component, are about $\Pra_T=7$ for the temperature and $\Pra_S=700$ for the salinity, respectively. Thus heat diffuses on a time scale two orders of magnitude faster than that of salinity. Due to this huge difference, long narrow vertical convection cells, which are called salt fingers, may develop even when the averaged fluid density is stably stratified~\citep{Stern:1960}. Salt finger is crucial to the salinity transfer~\citep{Turner:AnnRev:1974}. More generally and from now on, the component with the smaller diffusivity is called salinity, and the other one temperature.

Extensive studies have been conducted both experimentally and numerically to reveal the structure of salt fingers and the fluxes associated with them. Early experiments often focused on a single finger layer which grows from an interface of two homogeneous layers with different components. This includes the heat-salt system of \cite{Turner:1967}, \cite{McDougall_Taylor:1984}, and \cite{Taylor_Bucens:1989}, the heat-sugar system of \cite{Linden:DSR:1973}, and the sugar-salt system of \cite{Shirtcliffe_Turner:JFM:1970} and \cite{Pringle_Glass:JFM:2002}. Numerical simulations of DDC began in the 1980s and generated reasonable results compared to experiments~\citep[][and the references therein]{Yoshida_Nagashima:2003}. \cite{Sreenivas_etal:PoF:2009} conducted two-dimensional simulations of salt fingers starting from a sharp interface, and systematically investigated the relation between the control parameters and the finger width, vertical velocity and fluxes. In most of these studies, the salt fingers occupy more and more volume as they grow in height. Some large scale three-dimensional simulations have been performed for periodic domain with uniform background component gradient, such as \cite{Traxler_etal:JFM:2011,Stellmach_etal:JFM:2011,Mirouh_etal:AJ:2012,Wood_etal:AJ:2013}.

The DDC flow was also investigated for fluids bounded by two reservoirs with fixed temperature and salinity, such as \cite{Linden:1978,Krishnamurti:JFM:2003}. For different control parameters, single finger layer or alternating stacks of finger and convective layers was observed. The overall flux then depends on the number of flow layers between reservoirs. \cite{Hage_Tilgner:PoF:2010} (HT hereafter) conducted a series of DDC experiments with an copper-ion-concentration-heat system in an electrodeposition cell. For all the parameters they explored, one single finger layer emerges in the cell and is bounded by two thin boundary layers adjacent to the top and bottom walls. \cite{Schmitt:JMR:2011} performed a theoretical analysis to explain the finger-convection in the HT experiments, and \cite{Paparella_Hardenberg:PRL:2012} numerically simulated the DDC flow between two parallel free-slip plates for very large salinity Rayleigh number $Ra=10^{13}$.

One of the key issues of DDC flow is to understand the dependence of the fluxes on the control parameters. In early experiments, it has been found that the dimensional salinity flux follows a scaling law $F_S = C(\Delta S)^{4/3}$ with $\Delta S$ being the salinity difference across the finger layer and $C$ a function determined by experiment~\citep{Turner:1967, McDougall_Taylor:1984, Taylor_Bucens:1989}. The same scaling law was also obtained by~\cite{Radko_Stern:JFM:2000} using an asymptotic analysis. The experimental results of HT show good agreement with the $\propto(\Delta S)^{4/3}$ scaling, although the prefactor of the scaling law has to be determined by experiment. Recently, \cite{Radko_Smith:JFM:2012} proposed a model for double-diffusive transport with constant background gradients of temperature and salinity. The model predicts the heat and salt transport at a so-called equilibrium state, which happens when the growth rates of the primary and secondary instabilities are comparable. The growth rates are obtained by linear analysis for the primary instability and numerically for the secondary instability. And the ratio between two growth rates has to be determined by simulation data.  

In the field of Rayleigh-B\'{e}nard (RB) flow it is now widely accepted that there does not exist a single scaling exponent and the Grossmann-Lohse (GL) theory \citep{GL:JFM:2000, GL:PRL:2001, GL:PRE:2002, GL:PoF:2004, Stevens_etal:GLrefit} provides a unifying viewpoint for understanding the dependence of the heat flux on control parameters~\citep{Ahlers_etal:RMP:2009}. The model bases on the global balance between the dissipation rates and the convective fluxes of momentum and temperature. The predictions of the theory are in agreement with most of the experimental and numerical data \citep{Stevens_etal:GLrefit}. 

The purpose of the present study is twofold. First, we numerically simulate the DDC flow between two parallel plates, in a set-up which is the same as in HT. Direct comparison will be made between simulations and experiments for the same control parameters. Second, inspired by its success for RB flow, the GL theory will be applied to the DDC problem. As we will explain, we can in fact do so without introducing any new parameters, thus providing a new theoretical framework to understand the numerical and experimental data for DDC.

The structure of the paper is as follows: In Section 2 we will describe the theoretical formulation of the problem. In Section 3 we will provide the numerical setup and control parameters, along with the visualisation of salt fingers. Then we will show the system response to the control parameters in Section 4, and discuss the effects of the temperature field in Section 5, respectively. The application of the GL theory to DDC flow will be given in Section 6. Section 7 is left to conclusions.

\section{Governing equations}

We consider DDC flow between two parallel plates which are perpendicular to the direction of gravity and separated by a height $L$. The Oberbeck-Boussinesq approximation is employed, which means that the fluid density is assumed to depend linearly on the two scalar fields, namely temperature $T$ and salinity $S$,
\begin{equation}
  \rho(T, S) = \rho_0 [1 - \beta_T(T-T_0) + \beta_S(S-S_0)].
\end{equation}
Here $\rho_0$ is some reference density, and $\beta_T$ (and $\beta_S$) is the positive expansion coefficient associated to temperature (and salinity). The governing equations read~\citep{Landau_Lifshitz:FM,Hort_etal:PRA:1992}
\begin{subequations} \label{eq:ddc}
\begin{eqnarray}
  \partial_t u_i + u_j \partial_j u_i &=& 
       - \partial_i p + \nu \partial_j^2 u_i + g \delta_{i3} (\beta_T \theta - \beta_S s), \label{eq:mom}  \\[0.1cm]
  \partial_t \theta + u_j \partial_j \theta &=& 
         \lambda_T \partial_j^2 \theta 
         + \frac{\lambda_S k_T^2}{c_p T_0} \left[\frac{\partial \mu}{\partial s}\right]^0_{T,p} \partial_j^2 \theta 
         + \frac{\lambda_S k_T}{c_p} \left[\frac{\partial \mu}{\partial s}\right]^0_{T,p} \partial_j^2 s,  \label{eq:tem} \\[0.1cm]
  \partial_t s + u_j \partial_j s &=& \lambda_S \partial_j^2 s 
         + \frac{\lambda_S k_T}{T_0} \partial_j^2 \theta. \label{eq:sal}
\end{eqnarray}
\end{subequations}
The flow quantities include the velocity $\mathbf{u}(\mathbf{x}, t)$, the kinematic pressure $p(\mathbf{x}, t)$, the temperature field $\theta(\mathbf{x}, t)$, and the salinity field $s(\mathbf{x}, t)$. Both $\theta$ and $s$ are relative to some reference values, $g$ is the gravitational acceleration, $\nu$ is the kinematic viscosity, and $\lambda_T$ and $\lambda_S$ are the diffusivities of temperature and salinity. The last two terms of (\ref{eq:tem}) represent the Dufour effect, which is the heat flux driven by the salinity gradient. $c_p$ is the specific heat at constant $p$, $k_T$ the thermal diffusion ratio, and $\mu(T,s,p)$ the chemical potential. $\left[\partial \mu / \partial s\right]^0_{T,p}$ denotes the derivative of $\mu$ with respect to $s$ at constant $T$ and $p$. And the last term of (\ref{eq:sal}) denotes the Soret effect, which is the salinity flux driven by temperature gradient. The Soret effect is characterised by the separation ratio~\citep{Liu_Ahlers:PRE:1997} 
\begin{equation}
  \Psi = -\frac{\beta_S}{\beta_T}\frac{k_T}{T_0} 
       = -\frac{\beta_S}{\beta_T}S_0(1-S_0)S_T.
\end{equation}
Here $S_T$ is the Soret coefficient. The Dufour effect is characterised by $\Psi$, the Lewis number (often used in oceanography) $\Lew=\lambda_T/\lambda_S$ and the Dufour number~\citep{Hort_etal:PRA:1992}
\begin{equation}
  Q = \frac{T_0\beta_T^2}{c_p \beta_S^2}\left[\frac{\partial \mu}{\partial s}\right]^0_{T,p}.
\end{equation}
\cite{Hort_etal:PRA:1992} showed that, relative to the Fourier heat transfer, the magnitude of the second and third terms in the right hand side of (\ref{eq:tem}) are of order $\Lew^{-1}Q\Psi^2$ and $\Lew^{-1}Q|\Psi|$, respectively. For liquid mixture usually $\Lew^{-1}\sim10^{-2}$ and $Q\sim0.1$, and for gas mixture $\Lew^{-1}\sim1$ and $Q\sim10$. This implies that the Dufour effect in liquid mixtures can be $10^4$ times smaller than that in gas mixture. \cite{Liu_Ahlers:PRE:1997} measured the two coefficients $\Lew^{-1}Q\Psi^2$ and $\Lew^{-1}Q|\Psi|$ for several gas mixtures and for most cases they are smaller than $0.5$. Thus we can anticipate that the Dufour effect should have negligible effects in the present problem. The Soret effect may introduce new type of instabilities \citep{Turner:1985} and affect the onset of convection and pattern formation \citep{Cross_Hohenberg:1993, Liu_Ahlers:PRL:1996}. In the present paper we focus on the fully developed convection, and as in studies in the field we also neglect the Soret effect, as it is small for DDC of turbulent salty water. Then in equations (\ref{eq:tem}) and (\ref{eq:sal}) only the respective first term on the righthand side survives.      

The dynamical system (\ref{eq:ddc}) is constrained by the continuity equation $\partial_i u_i = 0$ and the appropriate boundary conditions. In the present paper, both the top and bottom plates are non-slip, i.e. $\mathbf{u}\equiv\mathbf{0}$. In the horizontal directions we use periodic conditions. The aspect ratio $\Gamma=d/L$, with $d$ being the domain width, indicates the domain size in the simulations. The dimensionless control parameters are the Prandtl numbers and the Rayleigh numbers of temperature and salinity, which are, respectively,
\begin{equation}
   \Pra_T = \frac{\nu}{\lambda_T},  \quad \Pra_S = \frac{\nu}{\lambda_S}, \quad 
   \Ray_T = \frac{g \beta_T L^3 \Delta_T}{\lambda_T \nu}, \quad 
   \Ray_S = \frac{g \beta_S L^3 \Delta_S}{\lambda_S \nu}.
\end{equation}
We define the total temperature or salinity difference as 
\begin{equation}
  \Delta_T = T_{bot} - T_{top}, \quad \Delta_S =  S_{top} - S_{bot},
\end{equation}
which ensures that the Rayleigh number is positive when the component destabilises the flow. Subscripts ``{\it top}\,'' and ``{\it bot}\,'' denote the values at the top and bottom plates, respectively. We note that $\Pra_S$ is also called the Schmidt number (\Sch). Other parameters can be calculated from the four numbers above. For instance, the Lewis number and the density ratio are
\begin{equation}
  \Lew = \lambda_T/\lambda_S = \Pra_S\,\Pra^{-1}_T, \quad
  R_\rho = (\beta_T \Delta_T)/(\beta_S \Delta_S) = \Lew\,\Ray_T\,\Ray^{-1}_S.
\end{equation}
The key responses of the system are the non-dimensional fluxes of heat and salinity and the Reynolds number\begin{equation}\label{eq:Nuss}
  \Nus_T = \frac{\langle u_3 \theta \rangle_A - \lambda_T\partial_3\langle\theta\rangle_A}{\lambda_T L^{-1} \Delta_T }, \quad
  \Nus_S = \frac{\langle u_3 s \rangle_A - \lambda_S\partial_3\langle s \rangle_A}{\lambda_S L^{-1} \Delta_S}, \quad
  \Rey = \frac{U_c L}{\nu}.
\end{equation}
Here $\langle\cdot\rangle_A$ denotes the average over any horizontal plane and time. Correspondingly, $\langle\cdot\rangle_V$ denotes the average over time and the entire domain. $U_c$ is a characteristic velocity.

Similar to RB flow, exact relations can be derived from (\ref{eq:ddc}) between the dissipation rates for momentum, temperature, and salinity and the global fluxes. It should be pointed out that these relations only hold provided that the cross-diffusion terms in (\ref{eq:tem}) and (\ref{eq:sal}) are negligible and the flow reaches a statistically steady state. Following~\cite{Shraiman_Siggia:PRA:1990}, one then readily obtains from the dynamical equations of $\theta^2$, $s^2$, and the total energy $u^2/2 - g\beta_T z \theta + g\beta_S z s$ the relations
\begin{subequations}\label{eq:dissip}
\begin{eqnarray}
   \epsilon_\theta &\equiv& \left\langle \lambda_T [\partial_i \theta]^2 \right\rangle_V  
                  = \lambda_T\, (\Delta_T)^2\, L^{-2}\, \Nus_T,   \label{eq:disst}  \\
   \epsilon_s &\equiv& \left\langle \lambda_S [\partial_i s]^2 \right\rangle_V 
                  = -\lambda_S\, (\Delta_S)^2\, L^{-2}\, \Nus_S,  \label{eq:disss} \\ 
   \epsilon_u &\equiv& \left\langle \nu[\partial_iu_j]^2 \right\rangle_V 
                  = \nu^3 L^{-4}\, \left[ \Ray_T\, \Pra^{-2}_T (\Nus_T - 1) 
                        - \Ray_S\, \Pra^{-2}_S (\Nus_S + 1) \right]. \label{eq:disse}
\end{eqnarray}
\end{subequations}
These exact relations are the cornerstones for applying the GL theory to DDC flow. Moreover, they can be used to validate the convergence of the simulation by checking the global balances between the dissipation rate and the flux, as we did in \cite{Stevens_etal:JFM:2010} for RB flow.

Above discussions provide several methods to calculate the Nusselt numbers. One can either compute $\Nus_T$ and $\Nus_S$ based on the definition (\ref{eq:Nuss}), in which the average can be taken as the surface averaging of the temperature and salinity gradients at the top or bottom plate, or by the volume average of the flux over the entire domain. At the same time, according to the exact relations (\ref{eq:disst}) and (\ref{eq:disss}) the Nusselt numbers can also be computed by the volume average of the dissipation rates. \cite{Stevens_etal:JFM:2010} have discussed in detail these four methods. The four methods must give identical values when the flow is fully resolved. This is used as a validation of the numerical setup.

\section{Numerical simulations and visualisations of salt fingers}

In our numerical simulation, equation (\ref{eq:ddc}) is non-dimensionalized by using the length $L$, the free fall velocity $U=\sqrt{g \beta_T |\Delta_T| L}$, and the temperature and salt concentration differences $|\Delta_T|$ and $|\Delta_S|$, respectively. Both the top and bottom plates are set to be no-slip and with fixed temperature and salinity. Here we always set $\Delta_T>0$ and $\Delta_S>0$. Thus the flow is driven by the salinity difference while it is stabilised by the temperature field. The computational domain has the same width in both horizontal directions and periodic boundary conditions are employed for the side walls. Similar to the experimental setup of HT, we start each case with a vertically linear distribution for temperature and uniform salinity equal to $(S_{top}+S_{bot})/2$, respectively. To trigger the flow motion, the initial fields are superposed with small random perturbations whose magnitudes are one percent of the corresponding characteristic values. The numerical scheme is the same as in \cite{Verzicco_Orlandi:1996} and \cite{Verzicco_Camussi:JFM:1999, Verzicco_Camussi:JFM:2003}. The salinity field is solved by the same method as for the temperature field. We use a double resolution technique to improve the efficiency. Namely, a base resolution is used for flow quantities except for the salinity field, which is simulated with a refined resolution. The details and validation of such method are reported in \cite{Ostilla_etal:2014}.

Two different types of simulations are conducted in the present work. For the first type we set the Prandtl numbers at $(\Pra_T,~\Pra_S)=(7,~700)$, which are the typical values of seawater. We vary $\Ray_S$ systematically for two temperature Rayleigh numbers $\Ray_T=10^5$ and $10^6$. The details of these simulations are summarised in table~\ref{tab1}. Moreover, in order to make a direct comparison to the experiments, five cases from HT are numerically simulated with exactly the same parameters, which are summarised in table~\ref{tab2}.  
\begin{table}
\renewcommand{\tabcolsep}{0.15cm}
\begin{center}
\begin{tabular}{ccccccccccc}
  $\Ray_T$ & $\Ray_S$ & $R_\rho$ & $\Gamma$ & $N_x (m_x)$ & $N_z (m_z)$ 
   & $\Nus_T$ & $\Nus_S$ & $\Rey$ & Dif$_T$ & Dif$_S$ \\[0.2cm]
  $1\times10^5$ & $1\times10^6$ & 10  & 2.5 & 192 (2) & 96 (2)  & 1.0052 & 8.6347 & 0.1107  & $<0.1\%$ & $0.50\%$ \\
  $1\times10^5$ & $2\times10^6$ & 5.0 & 2.5 & 192 (2) & 96 (2)  & 1.0125 & 11.064 & 0.1814  & $<0.1\%$ & $0.31\%$ \\
  $1\times10^5$ & $5\times10^6$ & 2.0 & 2.0 & 192 (3) & 144 (2) & 1.0350 & 15.050 & 0.3521  & $<0.1\%$ & $0.25\%$ \\
  $1\times10^5$ & $1\times10^7$ & 1.0 & 2.0 & 240 (2) & 144 (2) & 1.0775 & 17.854 & 0.5254  & $<0.1\%$ & $0.16\%$ \\
  $1\times10^5$ & $2\times10^7$ & 0.5 & 1.6 & 240 (3) & 144 (2) & 1.1706 & 22.107 & 0.8275  & $<0.1\%$ & $0.91\%$ \\
  $1\times10^5$ & $5\times10^7$ & 0.2 & 1.6 & 240 (3) & 192 (2) & 1.4265 & 29.259 & 1.4652  & $<0.1\%$ & $0.40\%$ \\
  $1\times10^5$ & $1\times10^8$ & 0.1 & 1.6 & 288 (3) & 144 (2) & 1.8826 & 35.342 & 2.3496  & $0.19\%$ & $0.70\%$ \\[0.2cm]
  $1\times10^6$ & $1\times10^7$ & 10  & 2.0 & 240 (2) & 120 (2) & 1.0116 & 17.352 & 0.2773  & $<0.1\%$ & $0.69\%$ \\ 
  $1\times10^6$ & $2\times10^7$ & 5.0 & 1.2 & 192 (2) & 144 (2) & 1.0277 & 22.037 & 0.4584  & $<0.1\%$ & $0.39\%$ \\
  $1\times10^6$ & $5\times10^7$ & 2.0 & 1.2 & 240 (2) & 192 (2) & 1.0789 & 29.542 & 0.8727  & $<0.1\%$ & $0.44\%$ \\
  $1\times10^6$ & $1\times10^8$ & 1.0 & 1.0 & 240 (2) & 192 (2) & 1.1791 & 35.516 & 1.3349  & $0.25\%$ & $0.59\%$ \\
  $1\times10^6$ & $2\times10^8$ & 0.5 & 1.0 & 288 (2) & 192 (2) & 1.3929 & 42.500 & 2.0749  & $0.17\%$ & $0.83\%$ \\
  $1\times10^6$ & $5\times10^8$ & 0.2 & 1.0 & 360 (2) & 240 (2) & 2.0197 & 56.184 & 3.8484  & $0.59\%$ & $1.3\%$  \\
  $1\times10^6$ & $1\times10^9$ & 0.1 & 1.0 & 384 (3) & 385 (2) & 3.0231 & 68.098 & 6.2142  & $0.51\%$ & $1.7\%$  \\
\end{tabular}
\end{center}
\caption{Summary of the simulations with $\Pra_T=7$ and $\Pra_S=700$. Columns from left to right are: Rayleigh numbers of temperature and salinity, the density ratio, the aspect ratio of domain, resolutions in the horizontal and vertical directions (with refinement coefficients for multiple resolutions), Nusselt numbers of temperature and salinity, the Reynolds number based on the rms value of velocity, and maximal difference between the Nusselt numbers computed by four methods. The meshes in the $y$-direction are the same as in the $x$-direction.}
\label{tab1}
\end{table}%
\begin{table}
\begin{center}
\begin{tabular}{cccccccccccccccc}
 $\Pra_T$ & $\Pra_S$ & \begin{tabular}{@{}c@{}}$\Ray_T$ \\ (x$10^5$)\end{tabular} 
   & \begin{tabular}{@{}c@{}}$\Ray_S$ \\ (x$10^8$)\end{tabular} & $R_\rho$
   & $\Gamma$ & $N_x (m_x)$ & $N_z (m_z)$ & $\Nus_T$ & $\Nus_S$ & $\Rey$ & Dif$_T$ & Dif$_S$ & $\Nus^e_S$ \\[0.2cm]
 8.8 & 2031.3 & $4.19$ & $5.85$ & 0.17 & 1.0 &
   360 (2) & 288 (2) & 1.40 & 60.21 & 1.725 & $0.6\%$ & $1.7\%$ & 37.8  \\
 8.8 & 2046.1 & $4.18$ & $8.78$ & 0.11 & 1.0 &
   360 (2) & 288 (2) & 1.65 & 65.98 & 2.216 & $0.8\%$ & $1.7\%$ & 60.6 \\
 8.8 & 2044.2 & $20.9$ & $8.41$ & 0.58 & 0.6 &
   288 (2) & 288 (2) & 1.19 & 67.44 & 1.515 & $0.2\%$ & $1.0\%$ & 51.6 \\
 9.2 & 2229.8 & $61.2$ & $33.3$ & 0.44 & 0.5 &
   288 (2) & 360 (2) & 1.36 & 100.3 & 2.624 & $0.2\%$ & $1.4\%$ & 91.4 \\
 9.4 & 2309.6 & $121$  & $147$  & 0.20 & 0.4 &
   288 (3) & 432 (2) & 2.07 & 153.3 & 5.537 & $0.7\%$ & $1.9\%$ & 141.0 \\
\end{tabular}
\end{center}
\caption{Summary of simulations of five experimental cases. The first two columns are the Prandtl numbers and the last column is the experimental measurement of $\Nus_S$, respectively. The other columns are the same as in table~\ref{tab1}. The complete experimental results of these cases can be found in HT.}
\label{tab2}
\end{table}

\begin{figure}
\centering
\includegraphics[width=13cm]{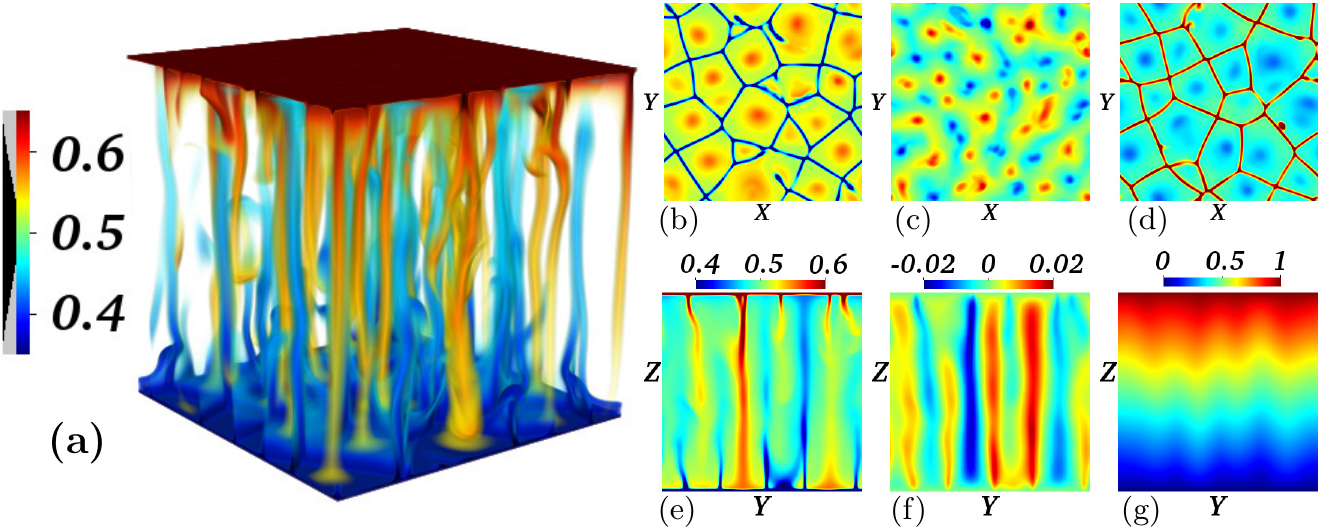}\\
\caption{Instantaneous flow visualisation for the case with $\Ray_T=1\times10^6$ and $\Ray_S=2\times10^8$. (a) Three-dimensional visualisation of salt fingers. Both the colour and the opacity are set by the salinity field. The red salty fingers grow from the top plate and extend to the bottom one, while the blue fresh fingers extend from the bottom plate to the upper one. (b-d) show the contours of $s$ on the horizontal planes with $z=0.05L$, $0.5L$, and $0.95L$, respectively. (e-f) show the contours on an vertical section plane  of salinity, vertical velocity, and temperature, respectively. (b-d) have the same colormap as (e).}
\label{fig1}
\end{figure}

In all simulations thin salt fingers grow from the boundary layers adjacent to both plates and extend through the entire cell height. Slender convection cells develop along with the salt fingers. In figure~\ref{fig1}(a) we show a three-dimensional visualisation of the salt fingers with $\Ray_T=10^6$ and $\Ray_S=2\times10^8$. The salty and fresh fingers locate alternatively in space and correspond to individual convection cells. Near top and bottom plates some sheet-like structures connects the roots of the fingers and form the boundaries of adjacent convection cells. 

To illustrate this more clearly, in figure~\ref{fig1}(b-d) we show salinity contours on three cross sections: $z=0.05$ near the bottom plate, $z=0.5$ in the middle plane, and $z=0.95$ near the top plate. Near both plates, the sheet-like structures are very distinct. The patterns are quite similar to those found in the sugar-salt experiments by \cite{Shirtcliffe_Turner:JFM:1970}. In the middle plane, the fingers take a nearly circular shape, although some weak links can be observed between fingers. This may explain why a ``sheet-finger'' assumption generates a better representation of the experimental data than a ``circular-finger'' in HT. These flow visualisations also suggest that the periodic condition in the horizontal directions is appropriate, provided that there are enough salt fingers and convection cells in the computational domain. For the case $(\Ray_T, \Ray_S)=(10^5,~5\times10^6)$ in table~\ref{tab1} we run a simulation with the same mesh size and half the domain size. The difference of the Nusselt numbers for the two domain sizes is smaller than $1\%$. For all simulations $\Gamma$ is so chosen that the flow domain contains similar amount of convection cells.

Figures~\ref{fig1}(e-f) depict the different patterns of the salinity, velocity, and temperature fields in a vertical plane. The salinity field has the smallest scale in the horizontal directions. Naturally, each salt finger is associated with a plume of high vertical velocity, which has a bigger width than salt finger. Due to its large diffusivity, the temperature field only exhibits wavy structures, and no thermal plumes can be found. The very different horizontal scales among various quantities verify the suitability and advantage of the double resolution method we used in our simulation. 

\begin{figure}
\centering
\includegraphics[width=10cm]{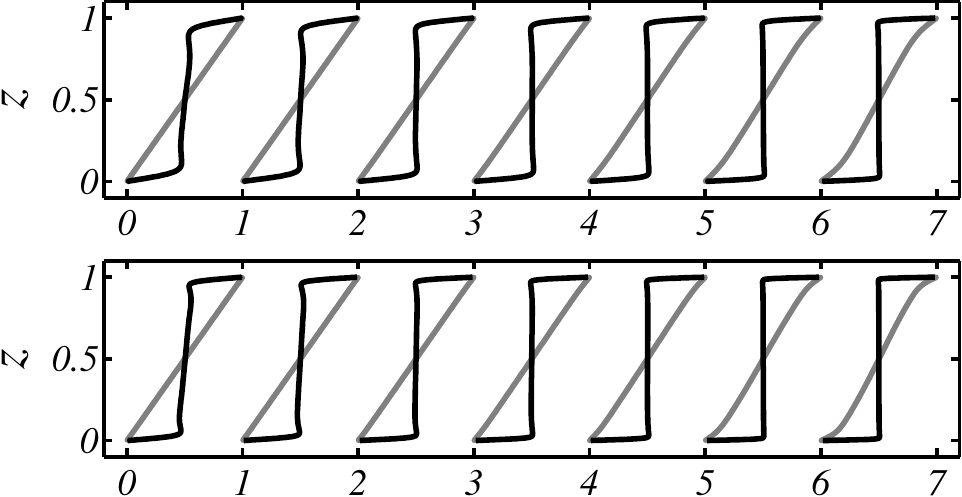}\\
\caption{Mean profiles of $s$ (black) and $\theta$ (grey) for cases in table~\ref{tab1}. Upper panel: $\Ray_T=10^5$ and from left to right $\Ray_S$ increases from $10^6$ to $10^8$; lower panel: $\Ray_T=10^6$ and from left to right $\Ray_S$ increases from $10^7$ to $10^9$, respectively. For clarity, curves are shifted rightward by $1$ from the previous one.}
\label{fig2}
\end{figure}

In figure~\ref{fig2} we plot the mean profiles $\overline{s}(z)$ and $\overline{\theta}(z)$ for the cases listed in table~\ref{tab1}. The overline stands for average over time and $(x, y)$ planes. Clearly, the salinity field has two distinct boundary layers adjacent to both plates, and in between there is a bulk region with $\overline{s}$ around $0.5$. As $\Ray_S$ increases, the thickness of the boundary layers decreases and the bulk region becomes more homogeneous. In contrast, there is no distinct division of boundary layers and bulk region in the temperature field. For given $\Ray_T$ when $\Ray_S$ is small the mean temperature profile keeps linear. Small deviations from the linear profile are only visible for large $\Ray_S$ (equivalently small $R_\rho$). This is reasonable since the fast diffusion of temperature (as compared to salinity) prevents the development of the small scale structures.

\section{System response in the explored parameter space}

\begin{figure}
\centering
\includegraphics[width=13cm]{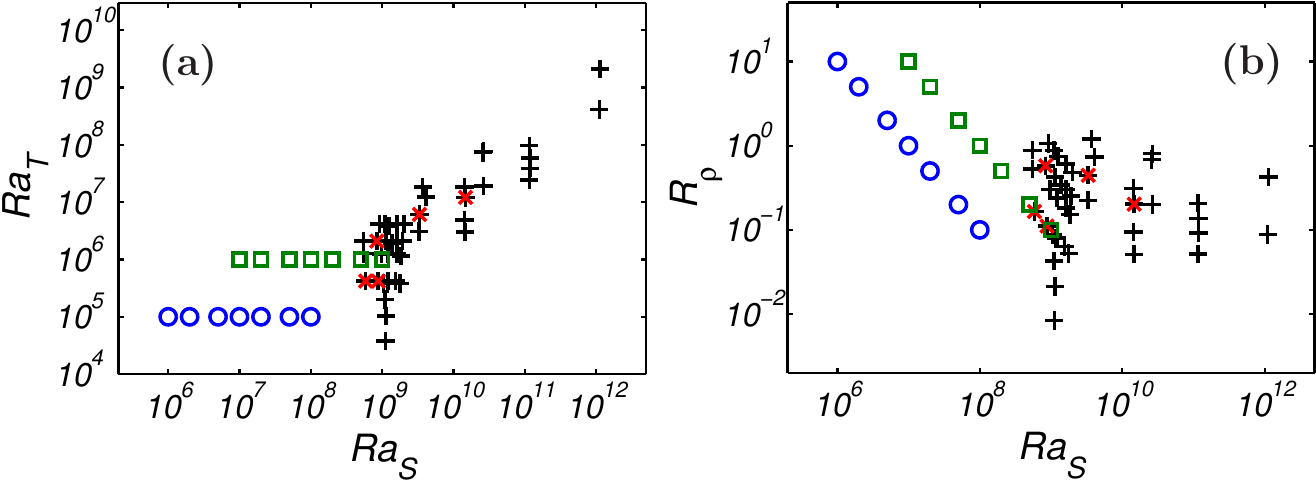}
\caption{parameter space in the $\Ray_S-\Ray_T$ and $\Ray_S-R_\rho$ planes. Blue circles: cases in table~\ref{tab1} with $\Ray_T=10^5$; green squares: cases in table~\ref{tab1} with $\Ray_T=10^6$; black pluses: experiments from HT; red crosses: cases listed in table~\ref{tab2}. The same symbols will be used in all the figure hereafter.}
\label{fig3}
\end{figure}

The parameter space we have explored is shown in the $(\Ray_S,\Ray_T)$ plane in figure~\ref{fig3}(a). The experimental cases of HT are also included in those figures. The cases listed in table~\ref{tab2}, which serve for the direct comparison between the numerical and experimental results, are marked by red crosses in the two parameter spaces. Together our simulations and HT experiments cover a $\Ray_S$ range over 6 decades and a $\Ray_T$ range over 4 decades, respectively. 

In figure~\ref{fig3}(b) we plot the same parameter space in the $(\Ray_S, R_\rho)$ plane. $R_\rho$ measures the ratio of the stabilising force of the temperature field to the destabilising force of the salinity field. When $R_\rho\ge1$, the flow is in the traditional finger regime which has been studied extensively. When $R_\rho<1$, the destabilising force of the salinity field is stronger than the stabilising force of the temperature field, thus the flow is more similar to RB flow. Most of the HT experiments are in the latter regime, and interestingly these authors found that fingers develop even with a very weak temperature difference. \cite{Schmitt:JMR:2011} extended the theory for the traditional finger regime to that with $R_\rho<1$ and revealed that narrow finger solution may still exist. Here in our simulation we systematically vary $R_\rho$ from $0.1$ to $10$, which covers both regimes.

As discussed in Section 2, the Nusselt numbers for temperature and salinity are measured by four different methods. The final Nusselt number is the average of these four values, which are given in tables~\ref{tab1} and \ref{tab2}. In those tables the maximal difference among the four values are also given. The maximal differences of $\Nus_T$ and $\Nus_S$ are less than $1\%$ for all the cases with $\Ray_S<5\times10^8$. When $\Ray_S\ge5\times10^8$ the differences increase but they are still below $2\%$. This confirms the convergence of our simulations. For Reynolds number we choose the characteristic velocity $U_c$ as the root mean square (rms) value of velocity computed with all three components.

\begin{figure}
\centering
\includegraphics[width=13cm]{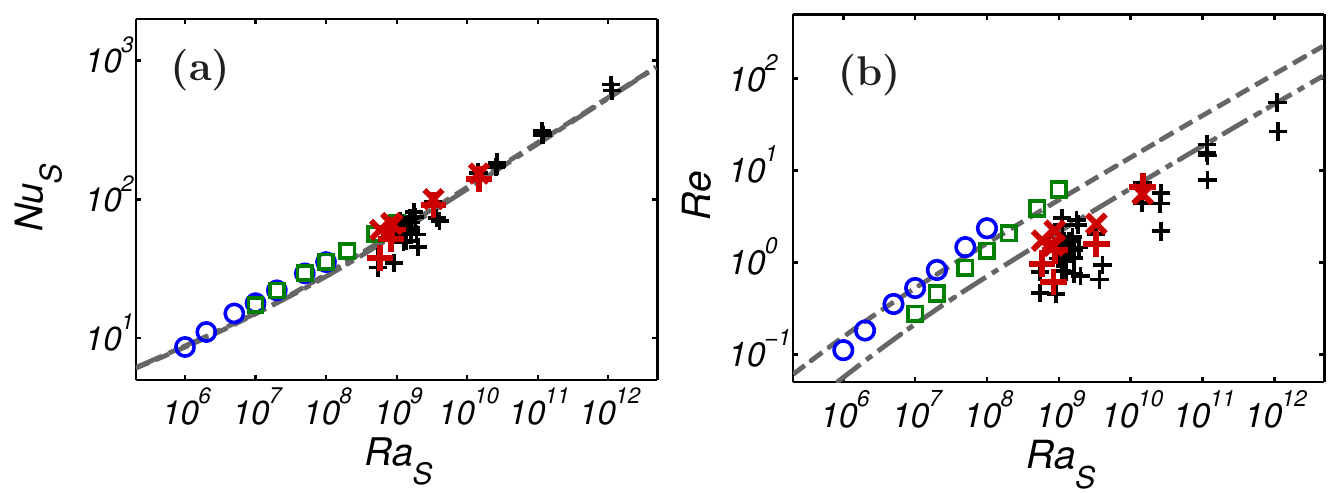}
\caption{$\Nus_S$ and $\Rey$ versus $\Ray_S$ for simulations and experiments. For the cases listed in table~\ref{tab2}, the red pluses represent the experimental results and the red crosses represent the simulation results. The predictions of GL model are given by the dashed line for $\Pra_S=700$ and the dash dotted line for $\Pra_S=2100$, respectively. For $\Nus_S$ these two lines almost collapse with each other (see (a)).}
\label{fig4}
\end{figure}

In figures~\ref{fig4}(a) and \ref{fig4}(b) we plot the dependences of $\Nus_S$ and $\Rey$ on $\Ray_S$, respectively. It is clear that $\Nus_S$ shows the same dependence on $\Ray_S$ in the whole range considered here, despite the different Prandtl numbers in simulations and experiments. In our simulations we have four pairs of cases at $\Ray_S=1\times10^7$, $2\times10^7$, $5\times10^7$, and $1\times10^8$. Within each pair $\Ray_T\in\{10^5, 10^6\}$ and we can see that $\Nus_S$ is very similar, \ie it has only a weak dependence on $\Ray_T$. Indeed, the symbols with same $\Ray_S$ and different $\Ray_T$ are very close to each other. Experimental results also show the same trend, especially in the higher $\Ray_S$ region. For instance, in figures~\ref{fig4}(a) at $\Ray_S\approx10^{11}$ there are actually four data points with $\Ray_T$ ranging from $2.42\times10^7$ to $9.7\times10^7$. This implies that $\Nus_S$ depends mainly on $\Ray_S$ and is only slightly affected by the change of $\Ray_T$.     

Changing $\Ray_T$ while keeping $\Ray_S$ fixed does have notable influence on $\Rey$, as shown in figure~\ref{fig4}(b). For same $\Ray_S$, larger $\Ray_T$ generates smaller $\Rey$. Recall that $\Ray_S$ measures the unstable driving force and $\Ray_T$ represents the stabilising force of temperature field. Then fixing $\Ray_S$ and increasing $\Ray_T$ means that the stabilising force becomes relatively stronger, therefore a smaller $\Rey$. The same phenomenon is also found in HT experiments.  

Cases in table~\ref{tab2} are marked by the red symbols in figures~\ref{fig4}. The red crosses are numerical results and the red plus marks are experimental results from HT. As compared to the experiments, the numerical simulations generate larger $\Nus_S$. It seems that the discrepancy becomes smaller as the experimental Nusselt number $\Nus^e_S$ increases. For the three cases with higher $\Nus^e_S$ the discrepancy is below $10\%$, which is within the uncertainty of experimental measurement. The discrepancy of $\Rey$ between experiments and simulations is larger than that of $\Nus_S$. This may be attributed to the way how the rms velocity is computed. HT computed the rms value by using the velocity components within a vertical plane where the flow field is measured. Here we compute the rms value by all three components and averaging over the entire domain. Since the salt fingers keep their position for very long time, the rms value measured in HT experiments depend on the location of the measured plane. Nonetheless, numerical results show a dependence of $\Rey$ on $\Ray_S$ similar to experiments. 

\section{Effects of the temperature field}

\begin{figure}
\centering
\includegraphics[width=8cm]{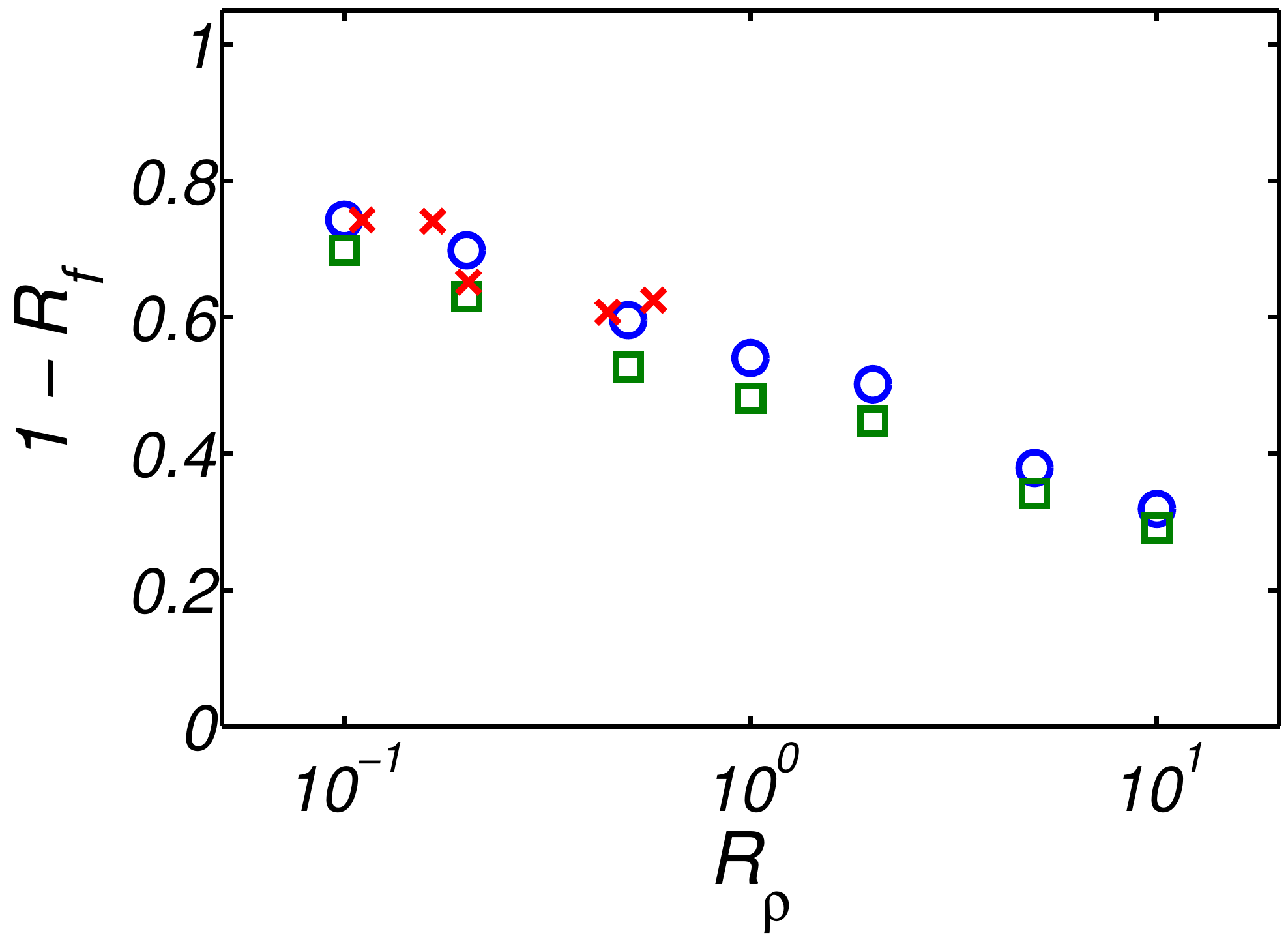}
\caption{Log-Linear plot of $1-R_f$ versus $R_\rho$. Blue circles: cases with $\Ray_T=10^5$ in table~\ref{tab1}; green squares: cases with $\Ray_T=10^6$ in table~\ref{tab1}; red crosses: cases in table~\ref{tab2}.}
\label{fig5}
\end{figure}

The density flux ratio, \ie the ratio of the density-anomaly fluxes due to temperature and that due to salinity, is defined as
\begin{equation}\label{eq:Rflx}
  R_f = \frac{\beta_T\langle u_3 \theta \rangle_V}{\beta_S\langle u_3 s \rangle_V}
      = \Lew R_\rho \frac{\Nus_T-1}{\Nus_S-1}.
\end{equation}
Then from (\ref{eq:disse}) one can easily get
\begin{equation}\label{eq:dspemodi}
  \epsilon_u = \nu^3 L^{-4} \Ray_S\, \Pra^{-2}_S (\Nus_S - 1) (1-R_f).
\end{equation}
Thus the temperature field affects the global balance between the momentum dissipation and the convection through the factor $1-R_f$. In figure~\ref{fig5} we plot the variation of $1-R_f$ versus $R_\rho$. Since the HT experiments did not measure the heat flux, in the figure we only show the numerical results. This dependence is similar for all analysed $\Ray_T$ (corresponding to different symbols in the figure). Namely, it decreases as $R_\rho$ increases. As $R_\rho\rightarrow0$, the influence of the temperature field becomes weaker and $R_\rho=0$ recovers the RB flow purely driven by the salinity difference. When $R_\rho\rightarrow\infty$, the stabilising force of the temperature field becomes stronger and eventually there is no motion. For $R_\rho=10$, $1-R_f$ is approximately $0.3$. Therefore the temperature field has a quite strong effect on momentum convection even when $\Nus_T$ is much smaller than $\Nus_S$. This is again due to the huge difference between $\lambda_T$ and $\lambda_S$, which is reflected by a large $\Lew$ in (\ref{eq:Rflx}). 

It should be pointed out that in our simulations $R_f$ increases as $R_\rho$ becomes larger, while it was reported in literature that $R_f$ is inversely proportional to $R_\rho$ for $R_\rho>1$ and $R_f \rightarrow 1$ as $R_\rho \rightarrow 1$, \eg see the review of~\cite{Kunze:PiO:2003} and the references therein. The reason for this difference may be the different flow configuration. For most of the experiments and simulations examined by~\cite{Kunze:PiO:2003} the fingers start from an interface between two homogeneous layers and grow freely during time. However, for our configuration the maximal vertical length of fingers is limited to the height between two plates and indeed for all the parameters we simulated fingers extend from one boundary layer to the opposite one and they have almost the same height.

\begin{figure}
\centering
\includegraphics[width=12.5cm]{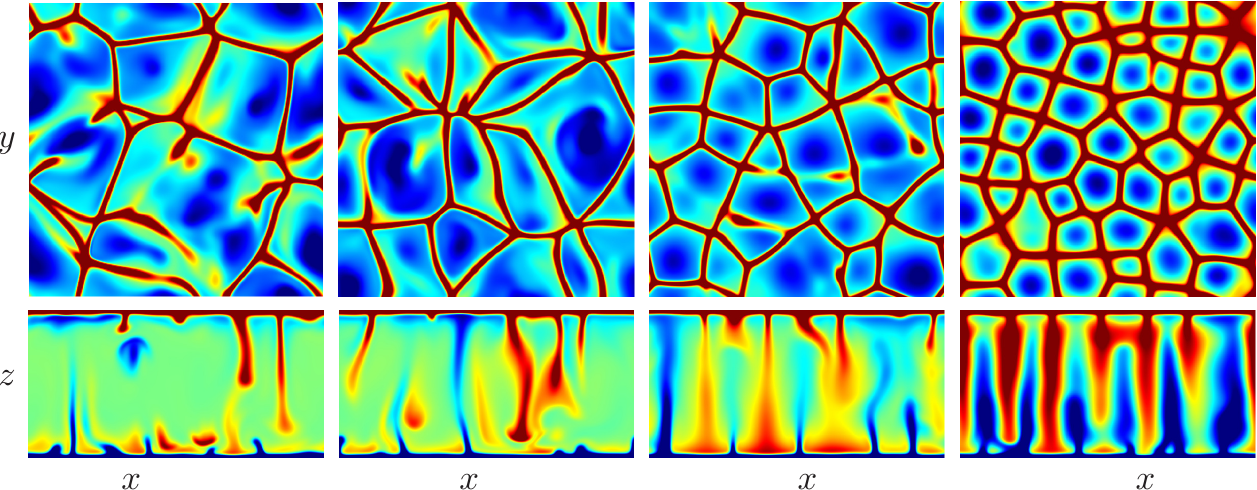}
\caption{Comparison of salinity fields for fixed $\Ray_S=10^7$ and different $\Ray_T$'s. From left to right: RB flow, $\Ray_T=10^3$, $\Ray_T=10^5$, and $\Ray_T=10^6$. Upper row: horizontal sections at $z=0.95$ near top plates; lower row: vertical sections. For all plots the colormap is the same as figure~\ref{fig1}(e).}
\label{fig6}
\end{figure}

To further reveal the effects of the temperature field, we simulated another two cases for $\Ray_S=10^7$. The first one has $\Ray_T=10^3$ and the other has no temperature field, namely pure RB flow. We obtain $\Nus_S=17.431$ and $17.249$, respectively. These values are very close to the two cases with the same $\Ray_S$ in table~\ref{tab1}, \ie $\Nus_S=17.854$ with $\Ray_T=10^5$ and $\Nus_S=17.352$ with $\Ray_T=10^6$. Figure~\ref{fig6} compares the salinity fields for these four cases with different $\Ray_T$. It can be seen that as $\Ray_T$ increases, the horizontal size of the convection cells shrinks. In the RB flow shown in the left column, the salt plumes from one plate becomes very weak before they reach the opposite plate. For larger $\Ray_T$ the plumes are stronger and grow more vertically. And finally almost all plumes reach the opposite plate and form salt fingers. Therefore, with the stabilising effect of the temperature field, the large scale flows in the pure RB case are prevented and the salt fingers tend to move vertically.    

The above observations imply that the temperature field does change the morphology of the salinity field, such as the horizontal size of the salt fingers and convection cells. Recall that $\Rey$ has a notable dependence on $\Ray_T$ while the dependence of $\Nus_S$ on $\Ray_T$ is very weak. Thus it seems that the temperature field affects the size of salt fingers and the speed of the flow motion in such a way that the salinity flux keeps fixed for certain $\Ray_S$. 

\section{The GL theory applied to double diffusive convection}

The GL theory developed by \cite{GL:JFM:2000, GL:PRL:2001, GL:PRE:2002, GL:PoF:2004} successfully accounts for the $\Ray$ and $\Pra$ dependence of $\Nus$ and $\Rey$ number for RB flow. The starting point of the theory are two exact relations for the kinetic and thermal energy-dissipation rates (the analogues of (\ref{eq:dissip})). The volume average of the dissipation rates are then divided into the contributions of the bulk region and of the boundary layers, which both can then be modelled individually, leading to
\begin{subequations} \label{old-gl} 
\begin{eqnarray}
 && (\Nus - 1) \Ray \Pra^{-2} = 
      c_1 \frac{\Rey^2}{g\left(\sqrt{\Rey_c/\Rey}\right)} + c_2 \Rey^3, \\[0.2cm]
 && \Nus - 1 = c_3 \Rey^{1/2}\,\Pra^{1/2} \left\{ f\left[ \frac{2a\Nus}{\sqrt{\Rey_c}}
             g\left( \sqrt{\frac{\Rey_c}{\Rey}} \right) \right] \right\}^{1/2} \nonumber\\[0.2cm]
 && \hspace{2cm}  + c_4 \Rey\, \Pra f\left[ \frac{2a \Nus }{\sqrt{\Rey_c}}
                    g\left( \sqrt{\frac{\Rey_c}{\Rey}} \right) \right],
\end{eqnarray}
\end{subequations}
with $\Rey_c = 4a^2$ as critical Reynolds number, describing the transition to the large $Pr$ regime~\citep{GL:PRE:2002}. The model has five coefficients, \ie $a$ and $c_i$ with $i=1,2,3,4$. Their values are  $c_1=8.05$, $c_2=1.38$, $c_3=0.487$, $c_4=0.0252$, and $a=0.922$, with which $Nu(Ra,Pr)$ and $Re(Ra,Pr)$ can very well be described~\citep{Stevens_etal:GLrefit}.

In generalization of this concept, now all three dissipation rates (\ref{eq:dissip}) are split into bulk and BL contributions as
\begin{subequations}
\begin{eqnarray}
 \epsilon_u & = & \epsilon_{u,BL} + \epsilon_{u,bulk}, \\
 \epsilon_\theta & = & \epsilon_{\theta,BL} + \epsilon_{\theta,bulk}, \\
 \epsilon_s & = & \epsilon_{s,BL} + \epsilon_{s,bulk}.
\end{eqnarray}
\end{subequations}
If both the temperature and salinity differences drive the flow, then it is natural to model both components in the same way, namely,
\begin{subequations}\label{dd-gl}
\begin{eqnarray}
 && (\Nus_S - 1) \Ray_S \Pra_S^{-2} + (\Nus_T - 1) \Ray_T \Pra_T^{-2}
           = c_1 \frac{\Rey^2}{g\left(\sqrt{\Rey_c/\Rey}\right)} + c_2 \Rey^3, \\[0.2cm]
 && \Nus_T  - 1 = c_{3,\theta}  \Rey^{1/2}\,\Pra_T ^{1/2} 
           \left\{ f\left[ \frac{2a \Nus_T  }{\sqrt{\Rey_c}}
           g\left( \sqrt{\frac{\Rey_c}{\Rey}} \right) \right] \right\}^{1/2} \nonumber\\[0.2cm]
 && \hspace{4cm}    + c_{4,\theta} \Rey\, \Pra_T f\left[ \frac{2a \Nus_T }{\sqrt{\Rey_c}}
                               g\left( \sqrt{\frac{\Rey_c}{\Rey}} \right) \right], \\ 
 && \Nus_S - 1 = c_{3,s} \Rey^{1/2}\,\Pra_S^{1/2} \left\{ f\left[ \frac{2a\Nus_S}{\sqrt{\Rey_c}}
               g\left( \sqrt{\frac{\Rey_c}{\Rey}} \right) \right] \right\}^{1/2} \nonumber\\[0.2cm]
 && \hspace{4cm}    + c_{4,s} \Rey\,\Pra_S f\left[ \frac{2a\Nus_S}{\sqrt{\Rey_c}}
                               g\left( \sqrt{\frac{\Rey_c}{\Rey}} \right) \right]. 
\end{eqnarray}
\end{subequations}
On first sight one may think that the 7 constants $c_1$, $c_2$, $c_{3,\theta}$, $c_{4,\theta}$, $c_{3,s}$, $c_{4,s}$, and $a$ would have to be obtained from a fit to experimental or numerical data. However, it is much easier in this case: They can be deduced from the {\it limiting cases} for which of course the same constants hold as in the general case: Imagine $Ra_S=0$, \ie only thermal driving. Then eqs.~(\ref{dd-gl}) reduce to eqs.~(\ref{old-gl}) with $c_{3,\theta} = c_3$ and $c_{4,\theta} = c_4$, i.e.,the known values. Next, imagine $Ra_T = 0$, \ie only salinity driving. Then the salinity field takes the role of the thermal field in the standard RB case and thus eqs.~(\ref{dd-gl}) again reduce to eqs.~(\ref{old-gl}), with $c_{3,s} = c_3$ and $c_{4,s} = c_4$, \ie again the known values! Moreover, as by construction of the model the prefactors do not depend on the control parameters $\Ray_T$, $\Ray_S$, $\Pra_T$, $\Pra_S$, these equalities not only hold in the limiting cases but throughout and we have in general
\begin{equation}
c_{3,\theta} = c_{3,s} = c_3 \quad \hbox{and} \quad c_{4,\theta} = c_{4,s} = c_4 
\end{equation}
with the {\it known} values for $c_3$ and $c_4$ and  also for $c_1$, $c_2$, and $a$~\citep{Stevens_etal:GLrefit}.

If the flow is driven by one component and stabilised by the other one, which is the case in the current study, the driving component can still be modelled in the same fashion as in (\ref{dd-gl}), but the other component must be modelled differently. However, as we discussed in the previous section, $\Ray_T$ only has a minor effect on the salinity transfer $\Nus_S$ in our problem. Moreover, the temperature field shows no clear distinction between the boundary layer region and the bulk region, as indicated by the temperature field in figure~\ref{fig1}(g) and the mean profiles in figure~\ref{fig2}. Thus one can neglect the thermal terms in model~(\ref{dd-gl}) and still obtain accurate predictions of the salinity transfer. 

With the original values of the coefficients $\Nus_S(\Ray_S)$ has been computed for $\Pra_S=700$ and $2100$, which are shown by lines in figure~\ref{fig4}(a). Indeed, the GL predictions agree excellently with both numerical and experimental results in the whole range of $10^6<\Ray_S<10^{12}$. The two lines with different $\Pra_S$ have only slight difference. This is similar to the RB flow where the Nusselt number saturates when the Prandtl number is large enough. We want to emphasize that no new parameters are introduced and the model developed for RB flow also works remarkably well for the present DDC flow.   

What about the dependence of the Reynolds numbers on the control parameters? As pointed out by~\cite{GL:PRE:2002}, the distribution $\Nus_S(\Ray_S, \Pra_S)$ is invariant under the transformation
\begin{equation}
  a \rightarrow \alpha^{1/2} a, \quad c_1 \rightarrow c_1/\alpha^2, \quad
  c_2 \rightarrow c_2/\alpha^3, \quad c_3 \rightarrow c_3/\alpha^{1/2}, \quad
  c_4 \rightarrow c_4/\alpha.
\end{equation}
Following the procedure of \cite{Stevens_etal:GLrefit}, we use one case to fix the transformation coefficient $\alpha$ and thus rescale the Reynolds number $\Rey(\Ray_S,\Pra_S)$ to the present flow. By using the Reynolds number of the case with $(\Ray_T, \Ray_S)=(10^5, 10^7)$, $\alpha$ is determined as 0.126. The GL prediction of $\Rey(\Ray_S)$ is then computed with the transformed coefficients for $\Pra_S=700$ and $2100$, which is shown in figure~\ref{fig4}(b). The theoretical lines show reasonable agreement with simulations and experiments. Since the effect of the temperature field is not included in the current model, thus the dependence of $\Rey$ on $\Ray_T$ is absent in the theoretical prediction.

Figure~\ref{fig4} demonstrates the success of our approach: As theoretically argued, it is indeed possible to apply the GL model {\it with the known parameters for RB flow} to DDC flow. The model not only captures the variation trends of $\Nus_S$ and $\Rey$, but also shows quantitively agreement with numerical and experimental data on the log-log plot. 
 
\begin{figure}
\centering
\includegraphics[width=13cm]{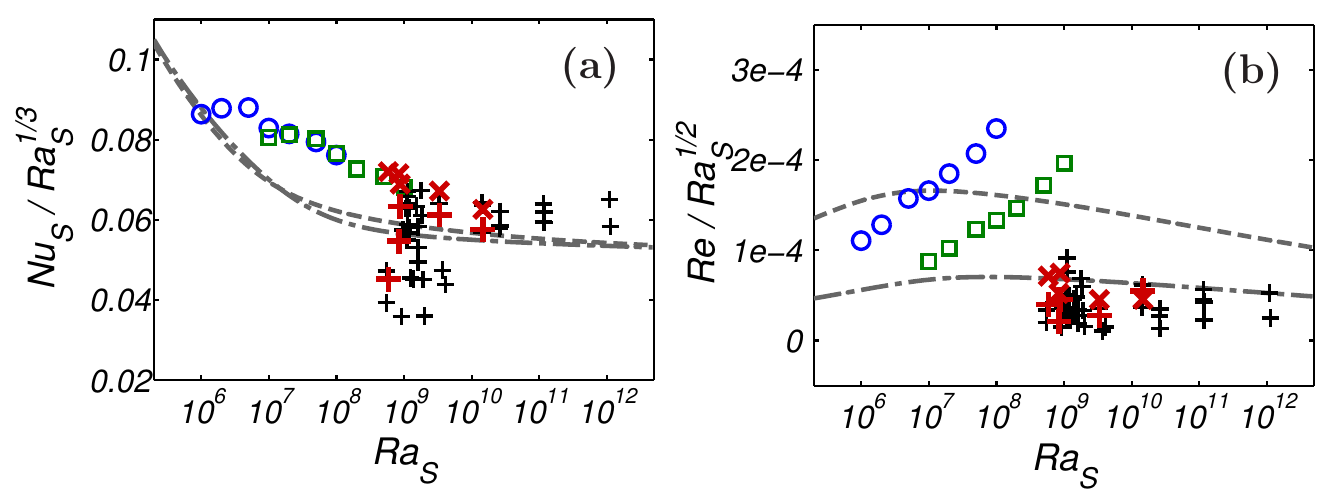}
\caption{Compensated plots of $\Nus_S$ and $\Rey$ versus $\Ray_S$. Lines and symbols are the same as in figure~\ref{fig4}.}
\label{fig7}
\end{figure}

In order to compare the model and the data more precisely, we plot the data and model predictions in a {\it compensated} way. Namely, $\Nus_S$ and $\Rey$ are respectively compensated by $\Ray_S^{-1/3}$ and by $\Ray_S^{-1/2}$. The results are shown in figure~\ref{fig7}. Here we see that our approach also inherits some weaknesses of the original GL model: Looking in this detail it becomes clear that $\Nus_S$ follows a trend different from the model, especially when $\Ray_S<10^9$. Similar discrepancy between the original GL model and experiment data was also observed for RB flow at very large Prandtl number, \eg see figure~7 of \cite{Stevens_etal:GLrefit}. The difference between the GL model and the data is even bigger for the Reynolds numbers of the cases in table~\ref{tab1}, as shown by the blue circles and dash line in figure~\ref{fig7}(b). Surprisingly, the model prediction of $\Rey$ shows a reasonable agreement with HT experiments even in the compensated plot.

As the end of this section, we would like to point out that the scaling laws given by \cite{Hage_Tilgner:PoF:2010}, which also capture the behaviour of the current numerical and experimental data, exhibit the similar transition around $\Ray_S\approx10^9$ when plotted in the compensated way as in figure~\ref{fig7} (not shown here). Therefore, the discrepancy at $\Ray_S<10^9$ requires further investigation in future work.

\section{Conclusions}

In conclusion, DDC flow was studied numerically for a series of flow parameters. Using a flow configuration similar to that of the experiments by~HT, in which the convection is driven by a salinity difference between two plates and stabilised by a temperature difference. Direct comparison was made between experiments and numerical simulations for several sets of parameters, and reasonable agreement is achieved for the salinity flux. Salt fingers exist in all the simulations. Flow visualisations show that the saltier and fresher fingers grow from the top and bottom plate, respectively, and extend to the opposite boundary layer. They are associated with slender convection cells. Near the plate where the saltier or fresher fingers grow, they usually originate from the sheet-like structures. When the fingers reach the opposite plate, they are bounded by the sheet-like structures near that plate. These sheet-like structures are quite weak in the bulk region. This justifies the ``sheet-finger'' assumption of~HT.    

Both our numerical results and experimental results of HT exhibit the same dependence of $\Nus_S$ on $\Ray_S$. For the present configuration, the change of $\Ray_T$ has minor influence on $\Nus_S$ but affects $\Rey$. To provide a new interpretation of the dependences of $\Nus_S$ and $\Rey$ on $\Pra_S$ and $\Ray_S$, we directly apply the Grossmann-Lohse theory for RB flow to the present problem. Without any modification of the coefficients, the theory successfully predicts $\Nus_S(\Pra_S, \Ray_S)$ with quite good accuracy for both numerical and experimental results in the $\Ray_S$ range of $(10^6, 10^{12})$. The $\Rey(\Pra_S, \Ray_S)$ prediction of the theory and the data also show reasonable agreement, especially for the experimental results. 

The effects of the temperature field are also discussed for the present flow configuration. The temperature field changes the morphology of the salt fingers but has minor influence on the salinity flux. It is remarkable that the Nusselt number of pure RB flow is very close to that of the double diffusive flow when the unstable component field has the same Prandtl and Rayleigh numbers.  

Finally, it should be pointed out that the current model does not include the influence of $\Ray_T$ and $\Pra_T$. Thus it cannot predict the behaviour of $\Nus_T$, neither can it describe the dependence of $\Rey$ on $\Ray_T$. Thus an extension of the GL model would be needed in future work to fully cooperate with the DDC problem. 

\bigskip 

This study is supported by FOM and the National Computing Facilities (NCF), both sponsored by NWO, and ERC.

\bibliographystyle{jfm}

\end{document}